\newcommand{\ra}{\;\raise1.0pt\hbox{$'$}\hskip-6pt\partial\;}
\newcommand{\lo}{\;\overline{\raise1.0pt\hbox{$'$}\hskip-6pt\partial}\;}

\newcommand{\red}{\textcolor[rgb]{1.00,0.00,0.00}}
\def \<{\langle}
\def \>{\rangle}

\documentclass[twocolumn,amsmath,amssymb,floatfix,superscriptaddress,showkeys]{revtex4}

\usepackage{graphicx,epsfig,natbib,color,times,bm,amsmath,multirow,slashbox}
\usepackage[ddmmyyyy,hhmmss]{datetime}
\usepackage[colorlinks]{hyperref}

\begin{document}
\title{Possible associated signal with GW150914 in the LIGO data}

\author{ Hao Liu} \email[mail to: ]{liuhao@nbi.dk}
\affiliation{Discovery Center, The Niels Bohr Institute, Blegdamsvej 17, DK-2100 Copenhagen, Denmark}
\affiliation{Key laboratory of Particle and Astrophysics, Institute of High Energy Physics, CAS, China}
\author{Andrew D. Jackson}
\affiliation{The Niels Bohr International Academy, The Niels Bohr Institute, Blegdamsvej 17, DK-2100 Copenhagen, Denmark}

\begin{abstract}  
We present a simple method for the identification of weak signals associated with gravitational wave events.
Its application reveals a signal with the same time lag as the GW150914 event in the released LIGO strain data
with a significance around $3.2\sigma$. This signal starts about 10 minutes before GW150914 and lasts for
about 45 minutes.  Subsequent tests suggest that this signal is likely to be due to external sources.
\end{abstract}


\maketitle

\section{Introduction} 
\label{sec:introduction}
The announcement by LIGO of the first observed gravitational wave (GW) event GW150914~\citep{LIGO PRL}
has opened a new era in astrophysics and generated considerable interest in the observation and identification
of associated signals.  Currently, attention has largely been focussed on electromagnetic signals, especially
gamma rays~\citep{Fermi-GW150914,AGILE-GW150914,XSL-GW150914}, that have the potential to confirm
both the existence and nature of GW150914.  In this work, we consider a rather different approach intended to
identify weaker signals in the LIGO strain data that have the same time lag as GW150914 itself.  The observation of
such associated signals is potentially useful in understanding the nature of the primary GW event..

\section{How to search for an associated signal} 
\label{sec:How to search for an associated GW signal}

At the time of the GW150914 event, the LIGO experiment had two running sites: Hanford (H) and Livingston (L).
Let the filtered strain data from them be $H(t)$ and $L(t)$, and let the arrival time delay between the two sites
be $\tau$ in the allowed range of  $-10\le\tau\le10$\,ms.  For convenience, we write the H/L strain data in time
range $(t_a, t_b)$ as $H_{t_a}^{t_b}$ and $L_{t_a}^{t_b}$, respectively.  The correlation coefficient between
the H/L strain data at time $t$ with delay $\tau$ and window width $w$ is
\begin{equation}\label{equ:running window corr}
C(t,\tau,w) = {\rm Corr}(H_{t+\tau}^{t+\tau+w},L_t^{t+w}).
\end{equation}
(The GW150914 signal arrived first at the Livingston site and reached the Hanford site approximately $6.9$\,ms
later~\citep{LIGO PRL}. Thus, Eq.~\ref{equ:running window corr} has been written so that $\tau$ is positive for
GW150914.)  Here, ${\rm Corr}(x,y)$ is the Pearson cross-correlation coefficient between two records $x$ and $y$
defined as
\begin{equation}\label{equ:corr}
{\rm Corr}(x,y) = \frac{\sum{(x-\overline{x})(y-\overline{y})}}{\sqrt{\sum{(x-\overline{x})^2}\cdot\sum{(y-\overline{y})^2}}},
\end{equation}
where the sums extend over all entries contained in the time interval considered and where $\overline{x}$ and
$\overline{y}$ are the corresponding average values of the entries in $x$ and $y$, respectively.  Note that this
cross-correlator is independent of the overall scales of $x$ and $y$ and of their average values.

Characterizing the GW150914 by the time of its effective start, $t_{gw}$, and its duration, $w=0.1$\,s
\footnote{In the LIGO work, GW150914 is assumed to have a duration of 0.2\,s, but the strongest part of the
signals is found in the later 0.1\,s.}, we define
\begin{equation}
\label{equ:corr for GW150914}
C_{gw}(\tau) = {\rm Corr}(H_{t_{gw}+\tau}^{t_{gw}+\tau+0.1\,{\rm s}},L_{t_{gw}}^{t_{gw}+0.1\,{\rm s}}).
\end{equation}
Evidently the strongest correlation and the largest magnitude of $C_{gw}(\tau)$ is expected for $\tau=6.9$\,ms.  This is
confirmed by Fig.~\ref{fig:cc vs tau}.

If there is a secondary signal associated with the GW150914 event in some time interval given by $(t_1,t_2)$,
we would expect to find that the corresponding correlator, $C(t,\tau,w)$ is similar to $C_{gw}(\tau)$ in that
they will both show a strong anti-correlation for the same value of $\tau=6.9$\,ms.  This expectation is independent
of both the strength of the associated signal and its shape.  If, however, the associated signal is weak, its presence can
be obscured by background noise.  Since the background noise in the H and L detectors is presumed to be uncorrelated,
noise can be suppressed by integrating $C(t,\tau,w)$ over the time stream.  Thus, keeping $w=0.1$\,s, we introduce
\begin{equation}\label{equ:ave corr over t}
D(\tau) = \frac{1}{t_2-t_1}\int_{t_1}^{t_2}C(t,\tau,0.1\,{\rm s}) dt.
\end{equation}
The net contribution of noise to $D(\tau)$ should vanish with increasing $(t_1 - t_2)$.  If both $(t_2 - t_1)$ and the
duration of the associated signal are sufficiently long, we expect that $D( \tau )$ will reveal a significant correlation
for $\tau \approx 6.9$\,ms even if the associated signal is weak.

The plot of $C_{gw}(\tau)$ in the upper left panel of Fig.~\ref{fig:cc vs tau} clearly shows the strong anti-correlation
at $\tau = 6.9$\,ms as expected.  It would be natural to think that this local property of the cross-correlator is its only
interesting feature.  This is not the case. In the absence of noise, any true GW signal would render the records H and L
identically except for a time shift.  The cross-correlator thus has a form that is similar to that of a convolution of this
signal with itself.  Using the fact that these records are real functions of time, it is elementary to show that the Fourier
transform of the cross-correlator is simply the absolute square of the Fourier transform of the record signal itself.
This quantity is immediately recognized as the power spectrum of the signal.  This has important consequences.
To the extent that two distinct events in the same record have the same cross-correlator (as a function of $\tau$),
we know that they necessarily have a common power spectrum.  Since the power spectrum contains no phase
information, it is not the case that the identity of two power spectra implies identical signals in the time domain.
Nevertheless, the observation of similarities between $C_{gw}(\tau)$ and $D(\tau)$ might suggest useful constraints
on the physical mechanisms that produce them.  it is useful to employ the cross-correlator once more to quantify this comparison.  Thus, we define
\begin{equation}\label{equ:similarity}
E(\tau_1,\tau_2) = {\rm Corr}(C_{gw}(\tau),D(\tau)) \ \ {\rm with}\ \ \tau\in(\tau_1,\tau_2).
\end{equation}
\section{Search for the associated signal and results} 
\label{sec:result_of_the_search}

Our search for associated signals is based on the publicly available 4096 second H/L strain data taken with a sampling
rate of 16384\,Hz rate~\citep{LIGO PRL,LIGO Tech,LigoData,LigoData2016}.  Data filtering was performed following
our previous work~\citep{PAH2016}.  This data set starts at the GPS time 1126257414.  Here it is more convenient to
use seconds from the beginning of this data set to mark the time.  We always exclude 30 seconds at both ends of the
data set to avoid the edge problem, and we always exclude the region $\pm 60$ seconds around the GW150914 event
(located at $\sim$2048 seconds) in order to avoid possible contamination of associated signals\footnote{In fact, exclusion of
$\pm 60$ seconds around the GW150914 event does not significant affect the results.}.  Since results were found to
be insensitive to the window width, we used the fixed value $w=0.1$\,s for convenience.

\subsection{Qualitative test with a simple initial guess} 
\label{sub:Search by initial guess}
\begin{figure*}[!htbp]
  \centering
  \includegraphics[width=0.45\textwidth]{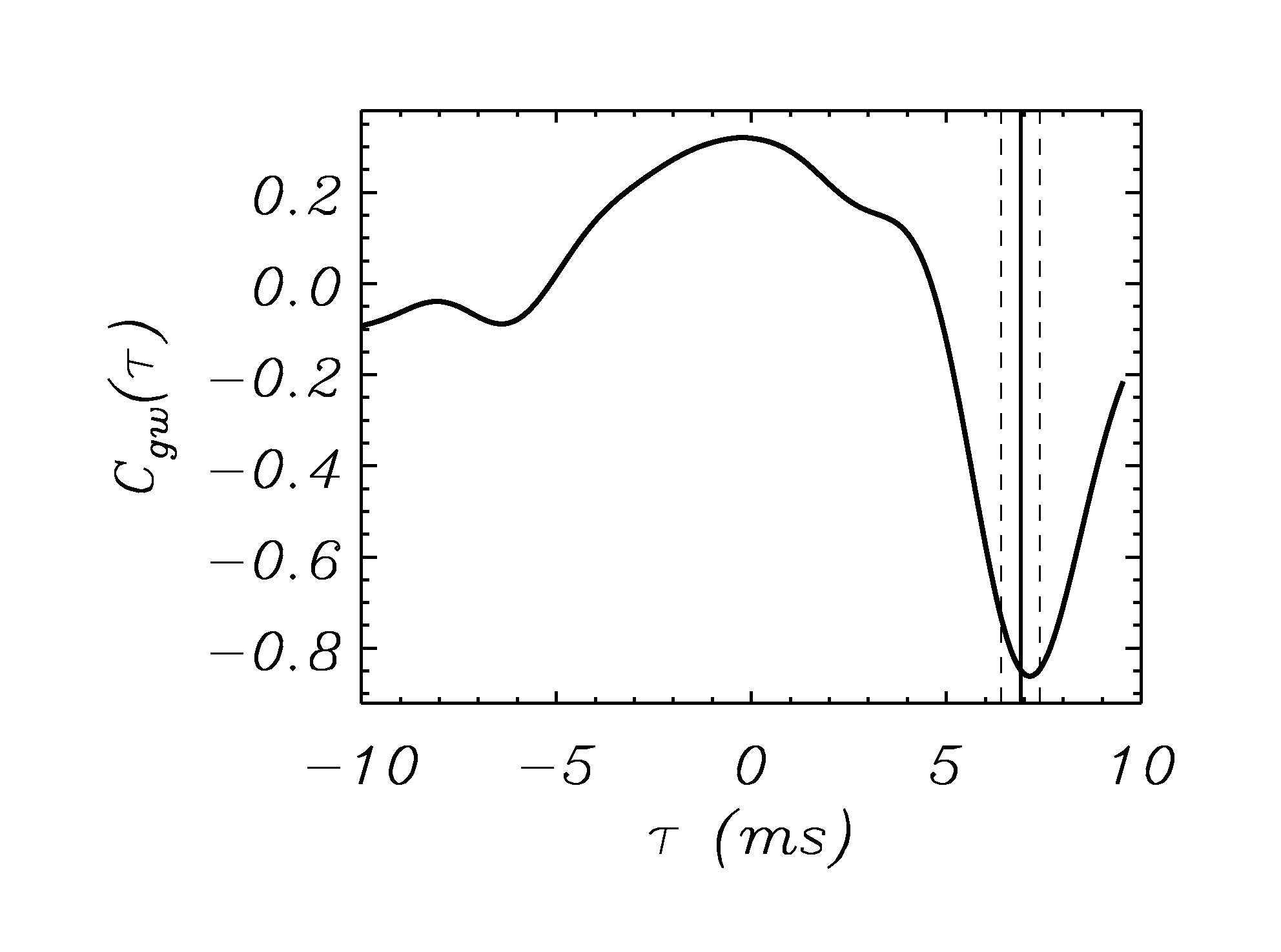}
  \includegraphics[width=0.45\textwidth]{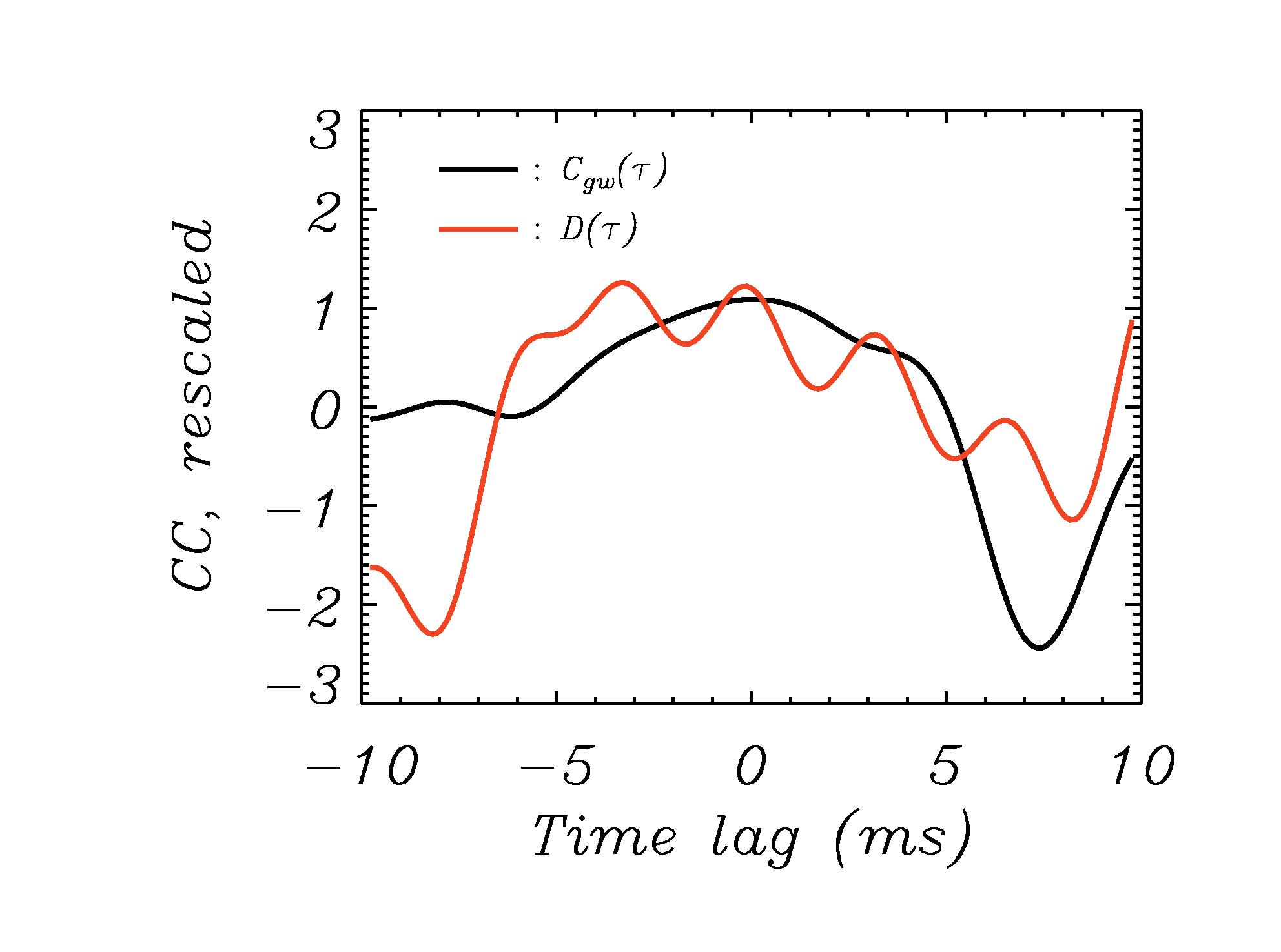} \\
  \includegraphics[width=0.45\textwidth]{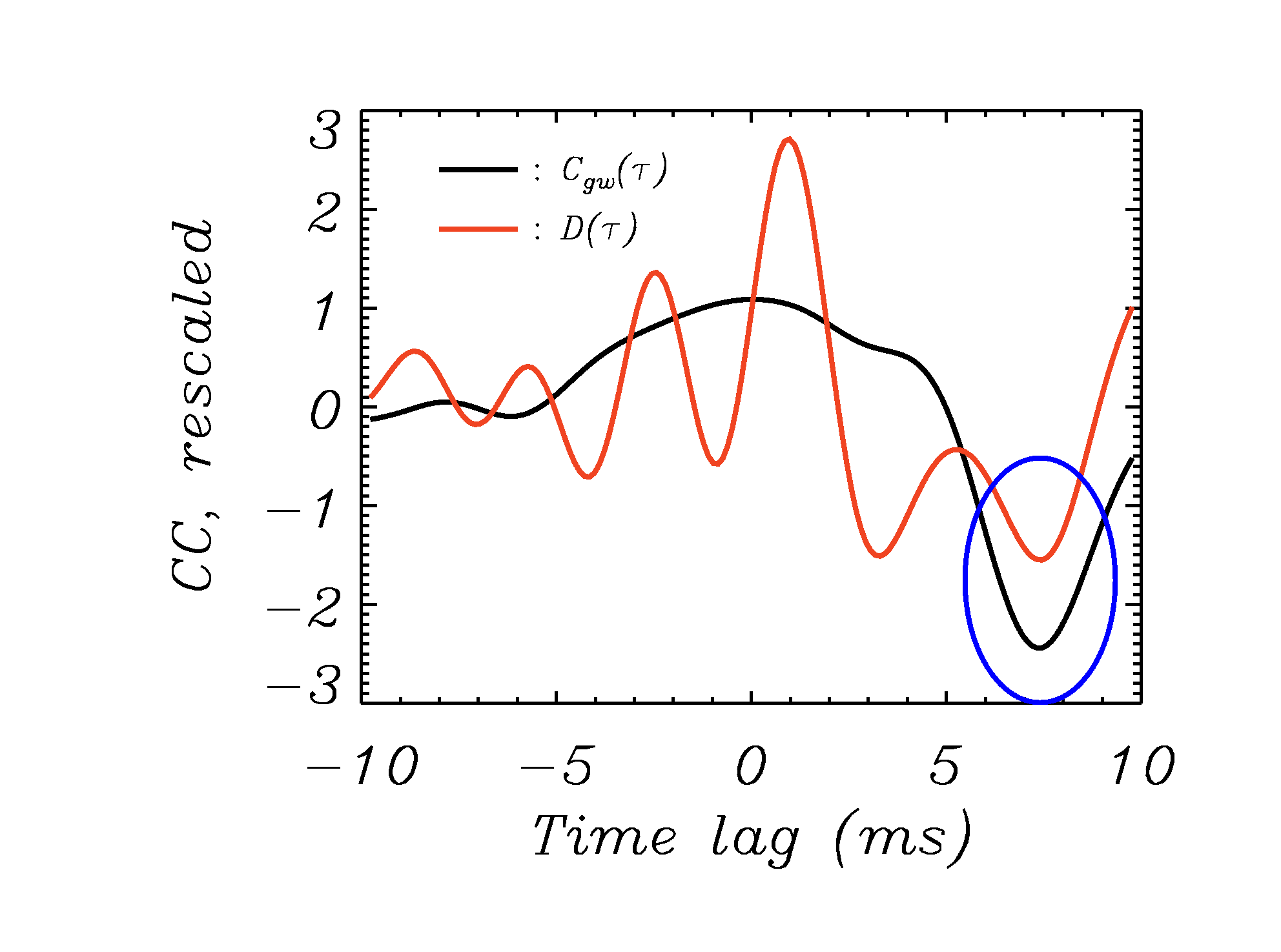}
  \includegraphics[width=0.45\textwidth]{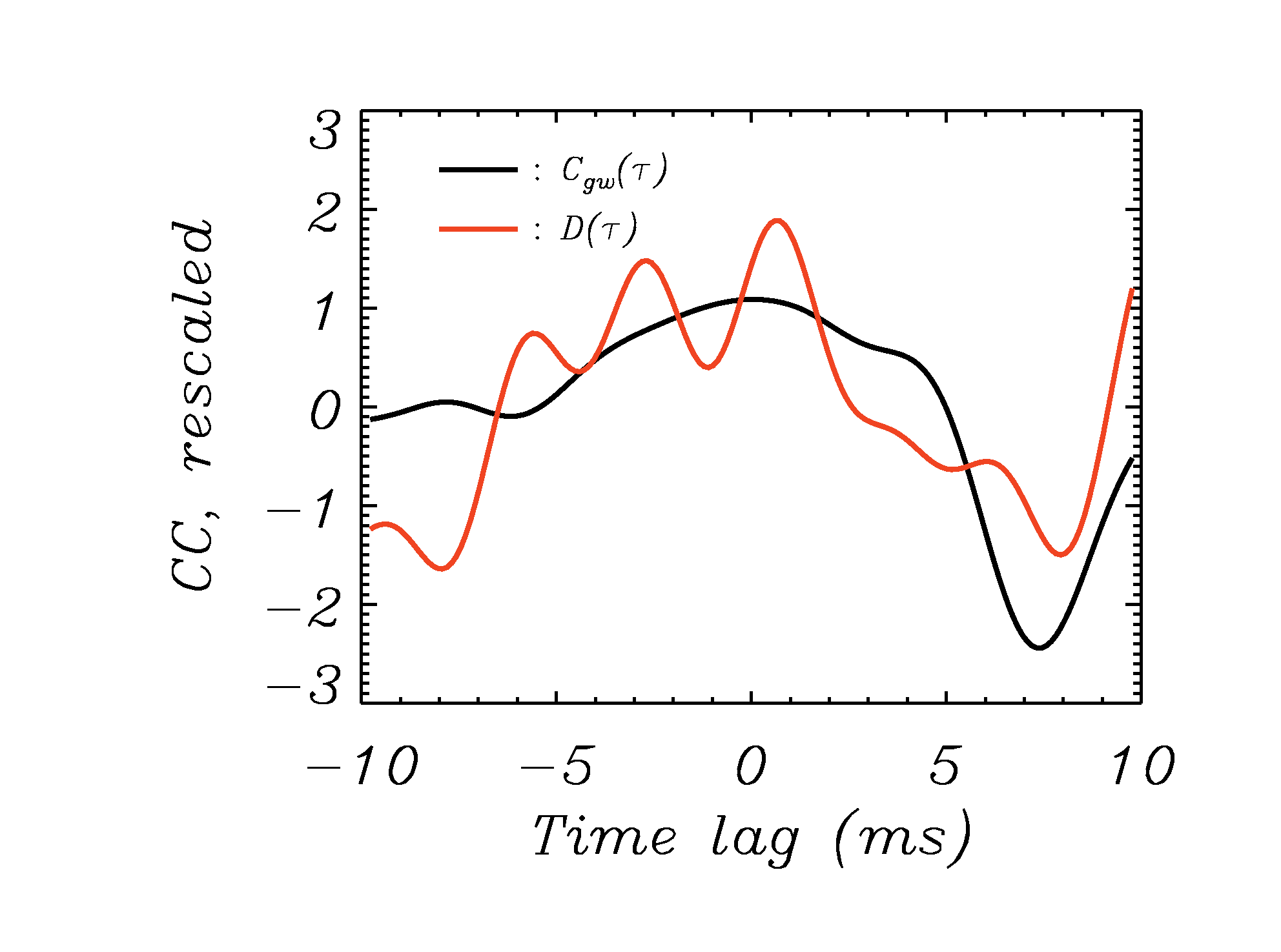}
  \caption{The H/L correlation vs. time lag for: GW150914 (upper-left), 30 minutes data before it, 30 minutes data after it
  (lower-left, see also Sec~\ref{sub:Search by initial guess}), and both (60 minutes, lower-right). The vertical lines indicate
  the $6.9\pm0.5$ ms range (official value of $\tau$ and its error for GW150914).   Data has been  rescaled as described in the text.}
  \label{fig:cc vs tau}
\end{figure*}

We begin by looking for a signal associated with GW150914 in an arbitrarily chosen region of $\pm 30$ minutes around the
primary event.  The resulting $D(\tau)$, defined in Eq.~\ref{equ:ave corr over t}, is shown in Fig.~\ref{fig:cc vs tau}.  The
upper right, lower left and lower right panels show data for 30 minutes before, 30 minutes after and 60 minutes surrounding
GW150914.  In these three panels we have shifted and rescaled both $D(\tau)$ and $C_{gw}(\tau)$ in such a way that
they each have an average value of zero and an rms value of 1.  This has been done to make the structure in $D( \tau )$
more visible.  As might be expected, $D(\tau)$ is quite small.  Consider, for example, the right panel of Fig.~\ref{fig:cc vs tau best range}, the average value of $D( \tau )$ before shifting is $-0.0002$, and the rms deviation from this average value is $0.0010$.  Although the resulting amplification of the cross-correlator is large, we shall demonstrate below that the structure shown in the figures is significant.  From this figure, we see that $D(\tau)$ shows its strongest correlation (marked with a
blue circle) for a value of $\tau$ consistent with the accepted value of $6.9 \pm 0.5$\,ms for GW150914.  Agreement is strongest for the data taken for 30 minutes after GW150914 (for which $E(5\,{\rm ms},9\,{\rm ms})$=0.82). These results and the rough similarity of $D(\tau)$ and $C_{gw}(\tau)$ over the entire physical range of $-10\,{\rm ms} \le \tau \le +10\,{\rm ms}$ encourage us to seek a more precise estimation of the time and duration of a possible associated signal.

\subsection{The duration of the associated signal} 
\label{sub:duration_of_the_associated_gw_signal}
\begin{figure*}[!htbp]
  \centering
  \includegraphics[width=0.33\textwidth]{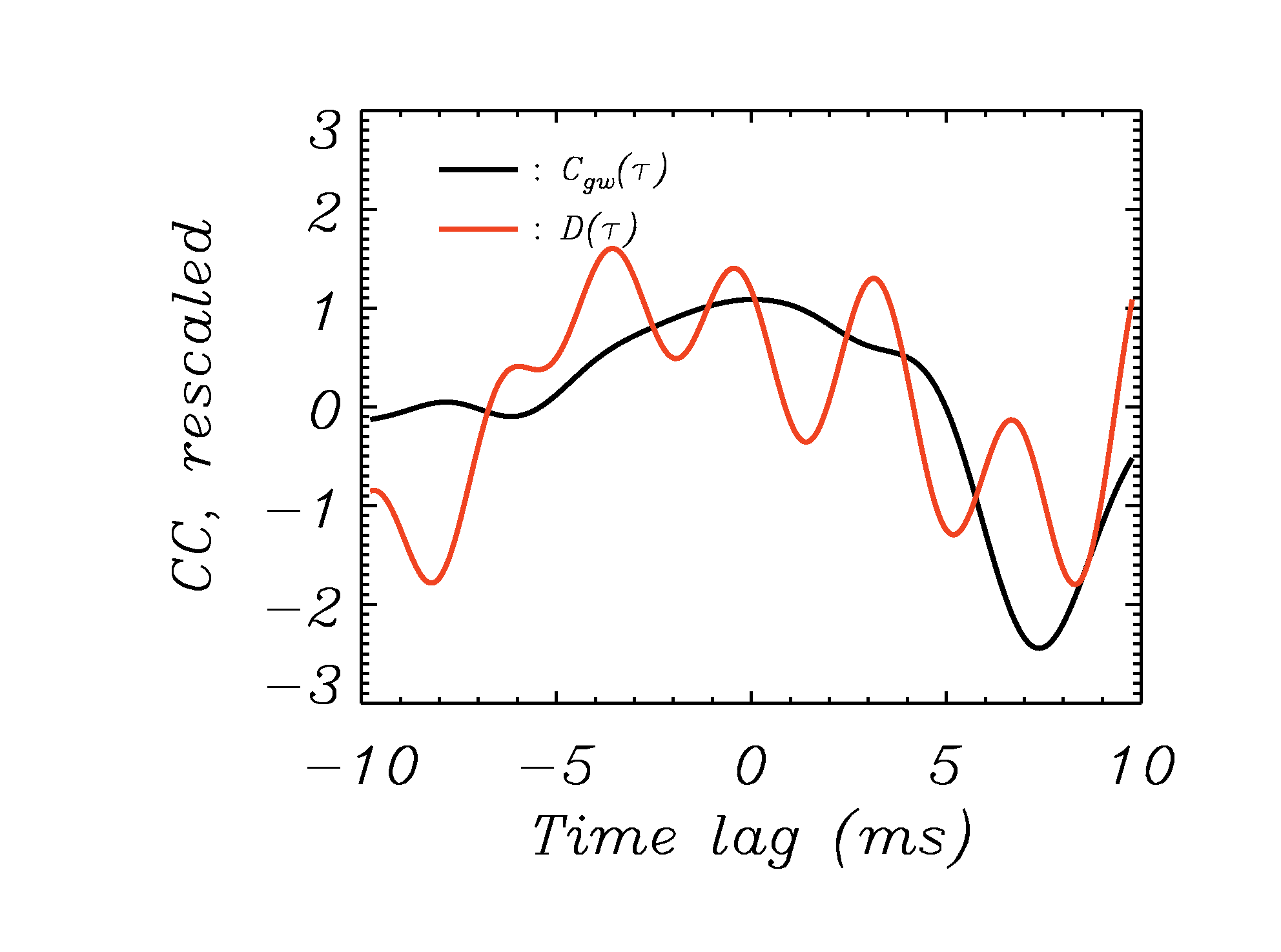}
  \includegraphics[width=0.33\textwidth]{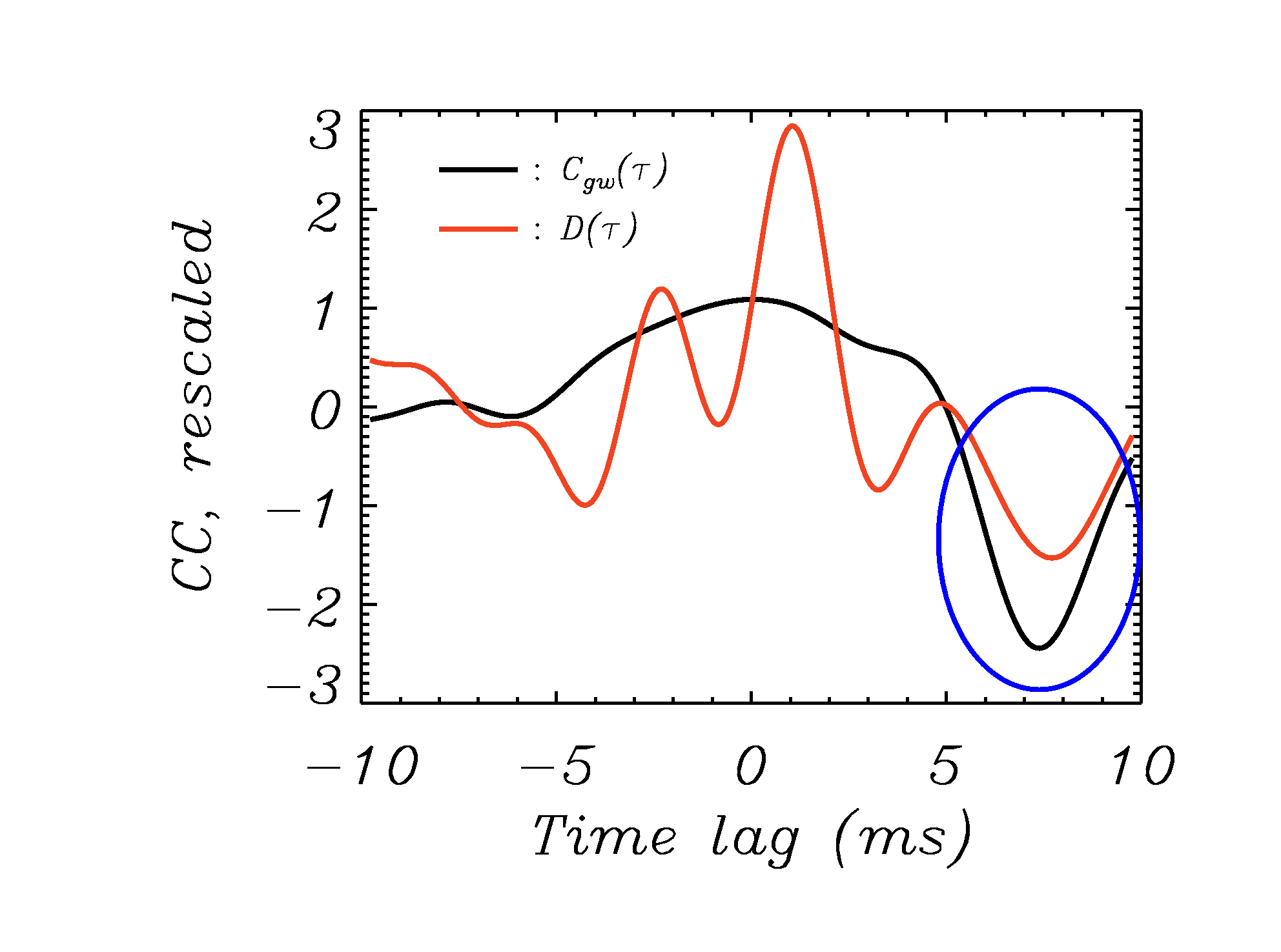}
  \includegraphics[width=0.33\textwidth]{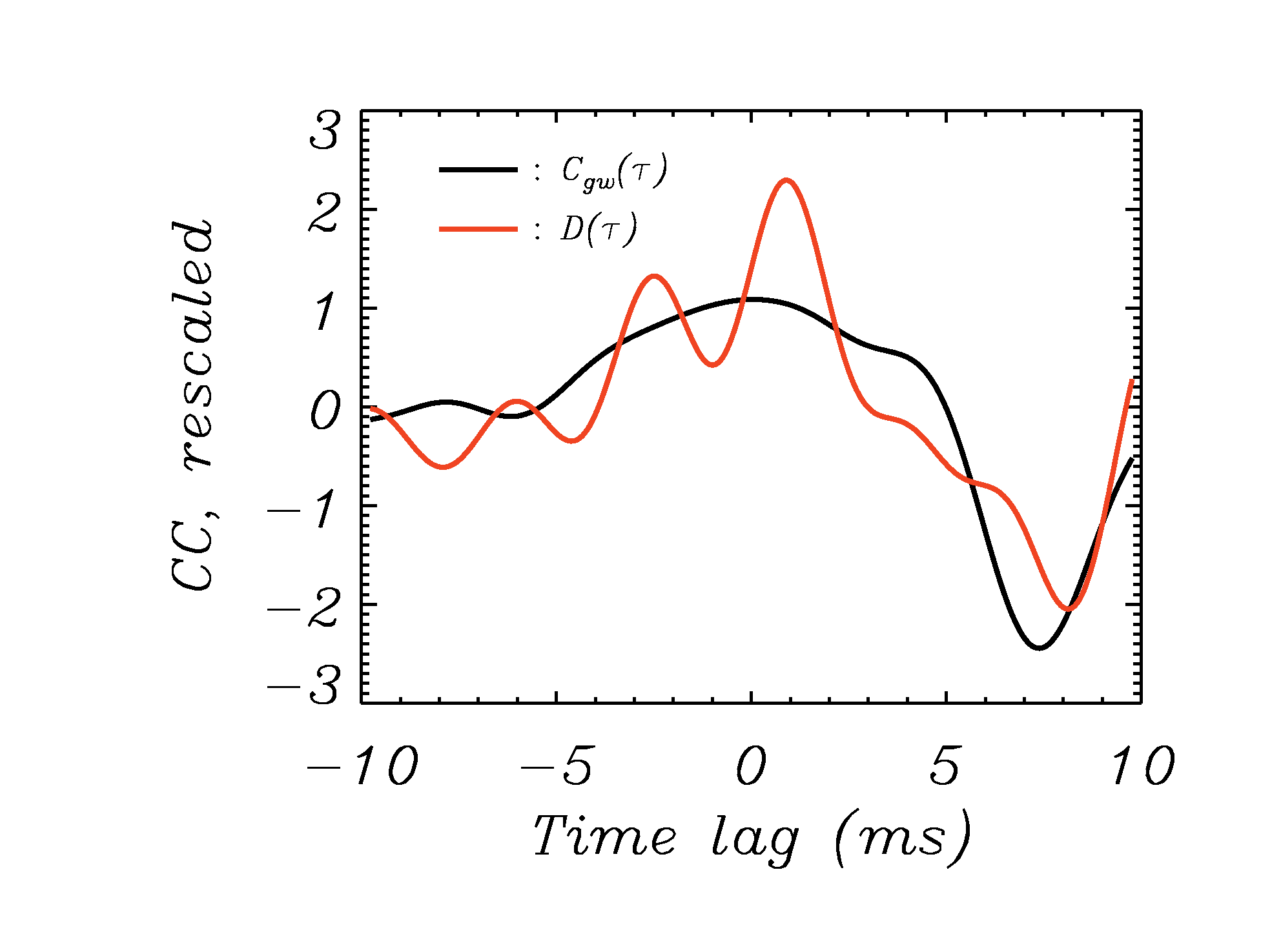}
  \caption{The H/L correlation versus time lag for \red{$(t_1,t_2)=(1280,4050)$} seconds. From left to right: before/after GW150914, and both.}
  \label{fig:cc vs tau best range}
\end{figure*}

In Sec~\ref{sub:Search by initial guess}, the time range was set arbitrarily at $\pm30$ minutes around GW150914.
One possible way of making an improved estimate of the actual duration of an associated signal is to turn to
Eq.~\ref{equ:ave corr over t}-\ref{equ:similarity} and adjust the interval $(t_1,t_2)$ in order to maximize
$E(-10\,{\rm ms},10\,{\rm ms})$. Doing this we find $t_1\approx1280$\,s and $t_2\approx4050$\,s, which means
that the associated signal starts $\approx$10 minutes before GW150914 and lasts for approximately
45 minutes.  The results obtained for $D(\tau)$ in this best-fit time range are shown in Fig.~\ref{fig:cc vs tau best range}.
The value $E(-10\,{\rm ms},10\,{\rm ms})$ in this time range is 0.84.  More details regarding $E(\tau_1,\tau_2)$ are given in Table~\ref{tab:cc tau best range}.

We note that the strongest correlation is again found close to $\tau = 6.9$\,ms as indicated by the blue circle in
Fig.~\ref{fig:cc vs tau best range}. The value $E(5\,{\rm ms},9\,{\rm ms})$ in the blue circle increases from 0.82
(Fig.~\ref{fig:cc vs tau}) to 0.96 (Fig.~\ref{fig:cc vs tau best range}).  Since the optimization was performed for
$\tau_2 - \tau_1 = 20$\,ms, this improvement provides some support for the validity of the estimation of
$(t_1,t_2)\approx(1280\,{\rm s}, 4050\,{\rm s})$.

\begin{table}[htbp]
 \centering
 \caption{Values of $E(\tau_1,\tau_2)$ for different intervals $(t_1,t_2)$. The three columns corresponds to the left,
 middle and right panels of Fig.~\ref{fig:cc vs tau best range}, respectively.}\begin{tabular}{|c|c|c|c|}\hline
                      & $(1280,2048)$ & $(2048,4050)$ & (1280,4050) \\ \hline
     $E(5\,{\rm ms},9\,{\rm ms})$ &     -0.21 &      \textbf{0.96} &      0.68 \\ \hline
  $E(-10\,{\rm ms},10\,{\rm ms})$ &      0.58 &      0.68 &      \textbf{0.84} \\ \hline
\end{tabular}
\label{tab:cc tau best range}
\end{table}

\subsection{The frequency range of the associated signal} 
\label{sub:About the frequency range of the associated GW signal}
\begin{figure}[!htbp]
  \centering
  \includegraphics[width=0.45\textwidth]{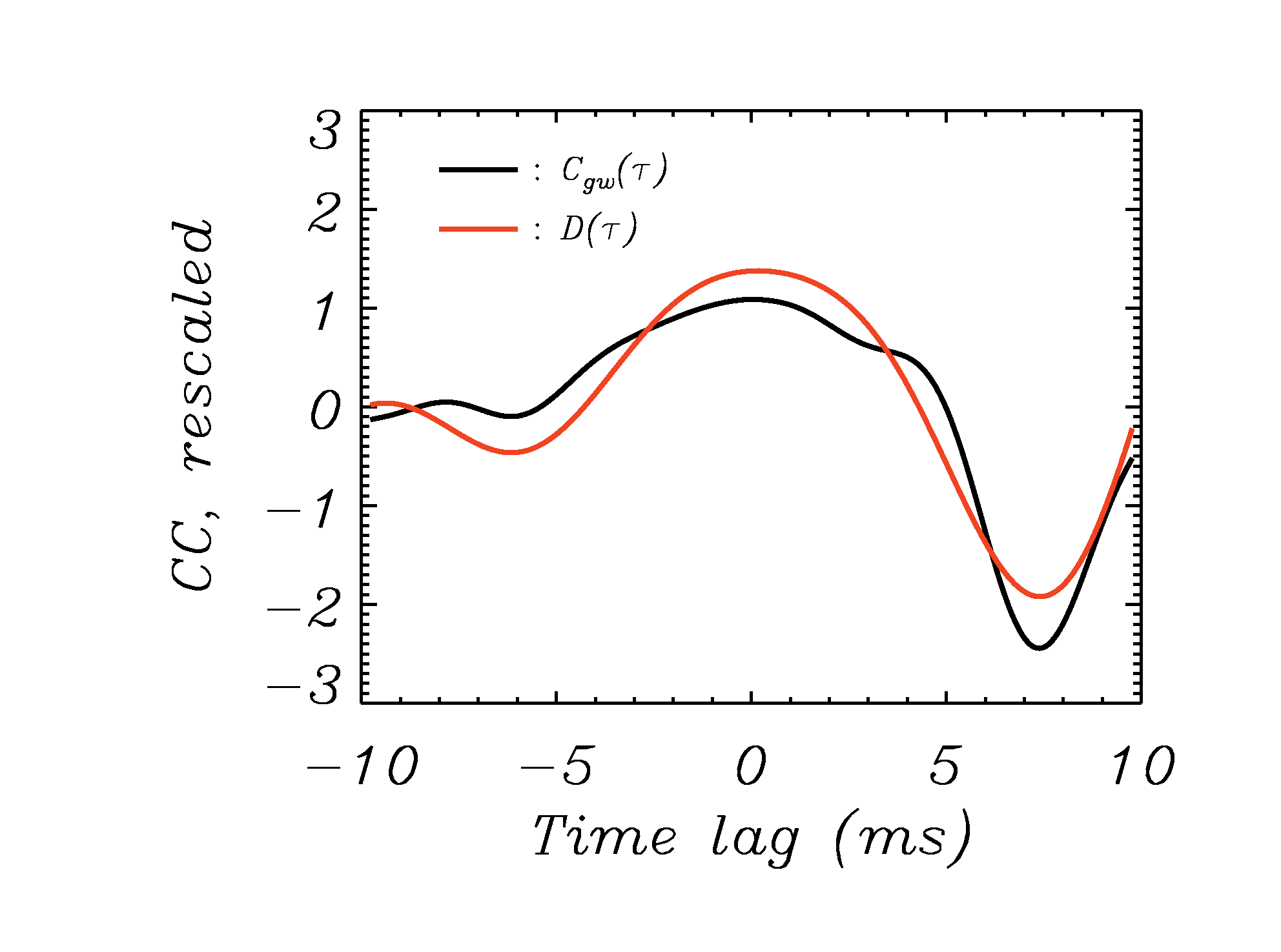}
  \caption{The H/L cross-correlator as a function of time lag for $(t_1,t_2)=(1280\,{\rm s},4050\,{\rm s})$ with a $50$--$150$\,Hz bandpass. The correlation between the black and red lines is $E(-10\,{\rm ms},10\,{\rm ms})=0.953$.}
  \label{fig:cc vs tau low freq}
\end{figure}

There is no a priori reason for the GW150914 event and its associated signal to have the same frequency range.  In particular, one might expect that higher frequencies might be less important in the far longer weak associated signal.  Thus, we have considered changing the bandpass range from $50$--$350$\,Hz to $50$--$150$\,Hz.  The result of this change is
that $E(-10\,{\rm ms},10\,{\rm ms})$ increases from 0.84 to 0.95 as shown in Fig.~\ref{fig:cc vs tau low freq}.
This remarkable strong similarity between $C_{gw}(\tau)$ and $D(\tau)$ (and hence their power spectra) would
appear to provide considerable support for the existence of an additional signal at lower frequencies associated with
GW150914.

\section{Tests and and the estimation of significance} 
\label{sec:testing_our_work}

\subsection{Possible systematic errors} 
\label{sub:self_consistency_test_ii_about_the_systematics_components}

Given the small amplitude of the associated signal, it is natural to consider the possibility that the similarity between
$C_{gw}(\tau)$ and $D(\tau)$ is due to unidentified systematic errors.  Here we partly reject this hypothesis by application
of increasingly narrow bandpass filters. More components of systematic error should be removed as we go through
this sequence of increasingly narrow bandpass filters.  The fact that this results in monotonically increasing values of
$E(-10\,{\rm ms},10\,{\rm ms})$ (Table~\ref{tab:cc tau freq range}) suggests that the correlation found in
Sec~\ref{sec:result_of_the_search} is not due to systematic errors.

\begin{table}[!htbp]
\caption{Values of $E(-10\,{\rm ms},10\,{\rm ms})$ for the time interval $(1280\,{\rm s},4050\,{\rm s})$ with increasingly
narrow band-pass filters. The final row (marked $*$) corresponds to the filtering adopted in Sec~\ref{sec:result_of_the_search}.}
\centering
\begin{tabular}{|c|r|}\hline
Bandpass range (Hz)     &  $E(-10,10)$    \\ \hline
All-pass                & -0.542          \\ \hline
$10\sim600$             &  0.003          \\ \hline
$20\sim550$             &  0.313          \\ \hline
$30\sim500$             &  0.680          \\ \hline
$40\sim350$             &  0.795          \\ \hline
\textbf{($*$)}          &  \textbf{0.843} \\ \hline
\end{tabular}
\label{tab:cc tau freq range}
\end{table}

Another test is also informative.  We split the frequency bands from 30 to 150\,Hz into six sub-bands, each of width
20\,Hz. In each sub-band we determine the value $E(-10\,{\rm ms},10\,{\rm ms})$.  The results are listed in
Table~\ref{tab:cc tau subband}. Here, we can see all six values of $E(-10\,{\rm ms},10\,{\rm ms})$ are positive.
Moreover, the correlations found in the $50-70$\,Hz and $90-110$\,Hz sub-bands are almost perfect.  The associated
signal appears to be broadly spread in frequency space.  Since LIGO systematic errors sources are normally of a well-defined frequency, the results of Table~\ref{tab:cc tau subband} suggest that the associated signal is more likely due to external sources than to systematic effects.

\begin{table}[!htbp]
\caption{Values of $E(-10\,{\rm ms},10\,{\rm ms})$ in sub-bands.}
\centering
\begin{tabular}{|c|c|c|c|c|c|c|}\hline
Band (Hz)    & 30-50 & 50-70 & 70-90 & 90-110 & 110-130 & 130-150 \\ \hline
$E(-10\,{\rm ms},10\,{\rm ms})$ & 0.67   & 0.98   & 0.25    & 0.99     & 0.47       & 0.31 \\  \hline
\end{tabular}
\label{tab:cc tau subband}
\end{table}

\subsection{Illustration using the LIGO S6 data} 
\label{sub:Test by using the S6 data}
In order to estimate the significance of the signal associated with GW150914, it would be useful to have a large data set that
does not contain any candidates for a strong GW event.  Unfortunately, no suitable data set is currently publicly available.
Thus, although it is far from ideal, we have considered the LIGO S6 data set~\citep{LigoDataS6}. This data, which predates
the LIGO upgrade that resulted in GW150914, was obtained under different physical conditions and may not be quantitatively
useful in understanding data associated with GW150914.  Therefore, for the purpose of illustration we downloaded S6 data
consisting of 300 records of 4096 duration each.  Each record was filtered with a 50$-$350\,Hz band-pass filter and proceed
precisely as above. It is clear that there can be remaining systematic effects or even artificial GW signals introduced by the
LIGO team for test purposes.  Ignoring all such concerns, we find that the value of $E(-10\,{\rm ms},10\,{\rm ms})=0.84$
for $(t_1,t_2)=(1280\,{\rm s},4050\,{\rm s})$ described above is larger than that the ones obtained from the S6 data set.
Thus the similarity between $C_{gw}(\tau)$ and $D(\tau)$ are likely not a consequence of noise or systematic effects.

\subsection{Significance estimation} 
\label{sub:Significance estimation}
\begin{figure}[htbp]
  \centering
  \includegraphics[width=0.45\textwidth]{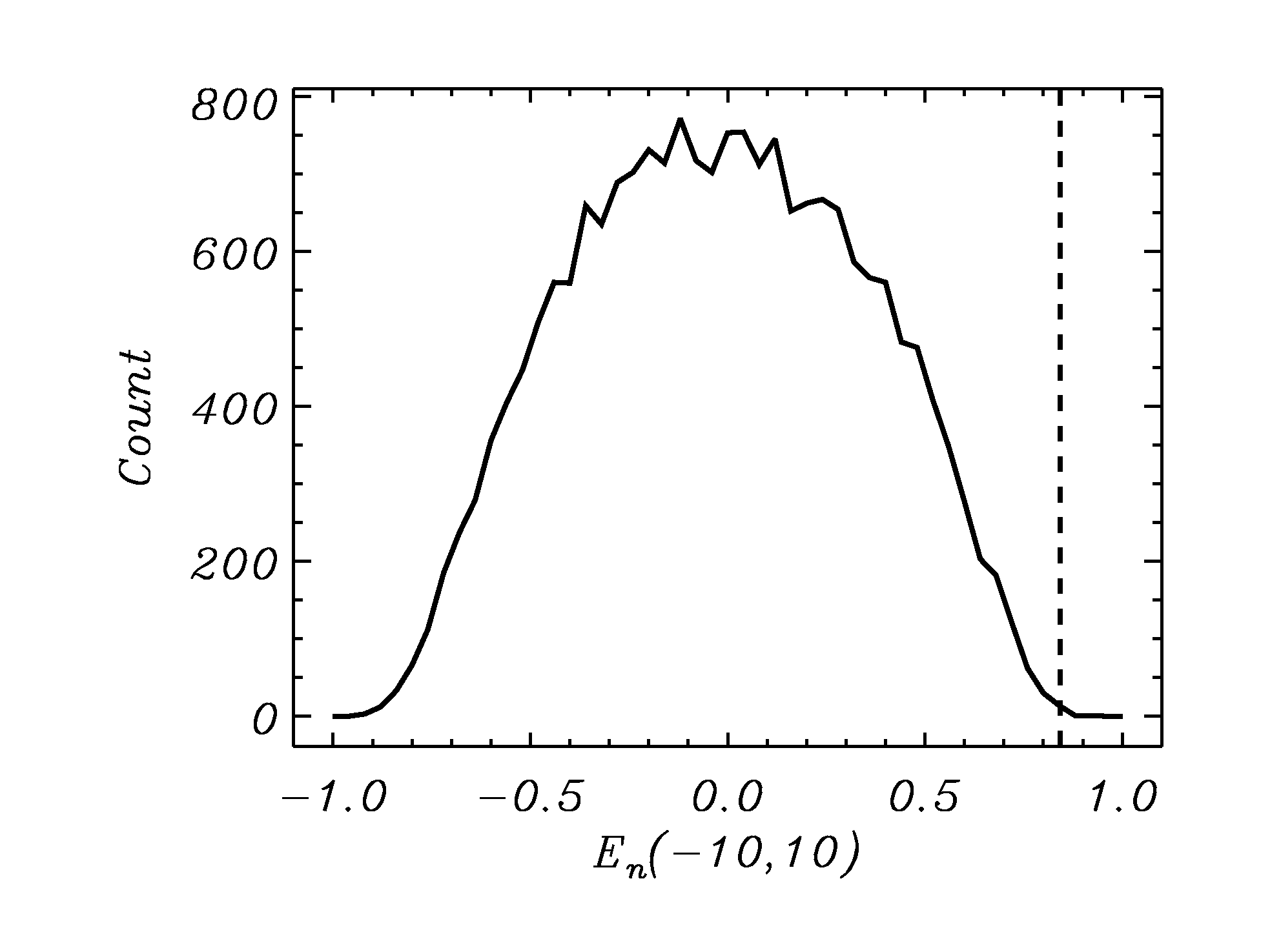}
  \caption{The histogram of 20,000 realizations of $E(-10\,{\rm ms},10\,{\rm ms})$ for noise as described in
  Sec~\ref{sub:Significance estimation}. The position of the real data is indicated by the dashed line.}
  \label{fig:hist_significance}
\end{figure}

Since the LIGO noise attribute is non-stationary and non-Gaussian, it is difficult to perform a reliable noise simulation.
In such cases, it is normal to use real data with an unphysical time lag (i.e., much larger than 10\,ms) in order to estimate the
significance of detection.  We employ a similar approach here.  Specifically, we take pairs of Hanford and Livingston data
segments from the filtered data that are 10--600 seconds away from each other in the time stream.  For each pair, we then
shift them again by $-10\sim10$\,ms in order to produce a realization of $D(\tau)$ for noise.  We then calculate
$E(-10\,{\rm ms},10\,{\rm ms})$ between each realization of $D(\tau)$ and the original $C_{gw}(\tau)$.  This process
was repeated 20,000 times.  The resulting distribution of cross-correlators is shown in the histogram of  Fig.~\ref{fig:hist_significance}.
For these 20,000 realization of the cross-correlator between $C_{gw}(\tau)$ and noise,
we find only 15 for which $E(-10\,{\rm ms},10\,{\rm ms}) > 0.84$.  This suggests that the probability of obtaining the
observed cross-correlation between $C_{gw}(\tau)$ and $D(\tau)$ accidentally and without a physical source is
$7.5\times 10^{-4}$, which corresponds to a significance of 3.2$\sigma$.

\section{Conclusion} 
\label{sec:Conclusion}
We have presented a method for identifying weak signals associated with gravitational wave events that is
based on time integrals of the Pearson cross-correlation coefficient.  Applying this method to the LIGO GW150914
event, we have found indications of an associated signal.  This signal has same time delay between Hanford and
Livingston detectors as the GW150914 event and has a duration of approximately 45 minutes (from approximately
12 minutes before to 33 minutes after GW150914).  Due to the weakness of this associated signal and its duration (which
appears to be 4 orders of magnitude greater than that of GW150914 itself), it is not possible to determine its shape in the
time domain. In spite of its weakness, however, we have argued that it is unlikely that this signal is of
systematic origin.  Numerical simulations show that this associated signal has a statistical significance of $3.2 \sigma$.  While it
is suggestive (but not conclusive) that this signal is real, it is not possible to offer a convincing suggestion regarding its
origin --- astrophysical or otherwise.  More generally, however, we have shown the value of studying the cross-correlation
between two identical signals as a function of the time shift, $\tau$, between them.  In such cases the Pearson cross-correlation
coefficient (as a function of $\tau$) is independent of the amplitude of the signals and directly related to the power spectrum
of the common signal.  Thus, the remarkable similarity of the cross-correlators found here for the associated signal
{\em before\/} GW150914, the GW150914 event itself, and the associated signal {\em after\/} GW150914 suggests that
their power spectra (but not their time domain shapes) are also similar.  Given the dramatic physical differences that are to be expected
in a system before, during and after a strong GW event, such a constraint on the associated power spectra could provide a
valuable diagnostic tool.

\begin{acknowledgments}
We would like to thank Pavel Naselsky, Alex Nielsen and Slava Mukhanov 
for valuable discussions.  This research has made use of data and software obtained from the LIGO Open Science Center
(https://losc.ligo.org), a service of LIGO Laboratory and the LIGO Scientific Collaboration.
LIGO is funded by the U.S. National Science Foundation.  This work was funded in part by the Danish National Research Foundation
(DNRF) and by Villum Fonden through the Deep Space project.  Hao Liu is supported by the National Natural Science Foundation for
Young Scientists of China (Grant No. 11203024) and the Youth Innovation Promotion Association, CAS.
\end{acknowledgments}

\end{document}